\documentstyle{mn}

\newif\ifAMStwofonts

\def\mincir{\raise -2.truept\hbox{\rlap{\hbox{$\sim$}}\raise5.truept \hbox{$<$}\ }}
\def\mincireq{\hbox{\raise0.5ex\hbox{$<\lower1.06ex\hbox{$\kern-1.07em{\sim}$}$}}}
\def\magcir{\raise-2.truept\hbox{\rlap{\hbox{$\sim$}}\raise5.truept \hbox{$>$}\ }}

\title{Photon propagation and the VHE $\gamma$-ray spectra of blazars: 
how transparent is really the Universe?}
\author[A. De~Angelis et al.]
       {A.~De~Angelis$^{1,2,3}$, O.~Mansutti$^{1,2}$, M.~Persic$^{3,2}$, 
        M.~Roncadelli$^4$ \\
        $^1$Dipartimento di Fisica, Universit\`a di Udine, via delle Scienze 208, I-33100 Udine, Italy \\
        $^2$INFN, Sezione di Trieste e Gruppo Collegato di Udine \\ 
        $^3$INAF, via G.B.Tiepolo 11, I-34143 Trieste, Italy \\
        $^4$INFN Pavia, Via A. Bassi 6, I-27100 Pavia, Italy} 
\date{Accepted ... ... ... ... .
      Received ... ... ... ... ;
      in original form ... ... ... ...}

\pubyear{2008}

\begin{document}

\maketitle

\label{firstpage}

\begin{abstract}
Recent findings by $\gamma$-ray Cherenkov telescopes suggest a higher 
transparency of the Universe to very-high-energy (VHE) photons than expected from 
current models of the Extragalactic Background Light. It has been shown that 
such transparency can be naturally explained by the DARMA scenario, in which the photon 
mixes with a new, very light, axion-like particle predicted by many extensions of the Standard 
Model of elementary particles. We discuss the implications of DARMA for the VHE 
$\gamma$-ray spectra of blazars, and show that it successfully accounts for the observed 
correlation between spectral slope and redshift by adopting for far-away sources the same 
emission spectrum characteristic of nearby ones. DARMA also predicts the observed blazar 
spectral index to become asymptotically independent of redshift for far-away sources. Our 
prediction can be tested with the satellite-borne {\it Fermi}/LAT detector as well as with the 
ground-based Cherenkov telescopes  H.E.S.S., MAGIC, CANGAROO\,III, VERITAS and the 
Extensive Air Shower arrays ARGO-YBJ and MILAGRO.
\end{abstract}


An impressive amount of information about the Universe at $\gamma$-ray energies 
larger than $100 \, {\rm GeV}$ -- namely in the very-high-energy (VHE) $\gamma$-ray band 
-- has been collected over the past few years by the Imaging Atmospheric Cherenkov 
Telescopes (IACTs) such as H.E.S.S., MAGIC, CANGAROO III and VERITAS. These IACTs 
have detected gamma-ray sources over an extremely wide interval of distances, ranging 
from the parsec scale for Galactic objects up to the Gpc scale for the farthest blazar 
3C\,279 at redshift $z = 0.536$ (Albert et al. 2008). 

These observations allow both to infer the intrinsic properties of the sources and 
to probe the nature of photon propagation through cosmic distances. The latter fact 
becomes particularly important in connection with VHE $\gamma$-ray observations, 
since hard photons travelling through cosmological distances interact with soft 
background photons permeating the Universe, producing  $e^+ e^-$ pairs through the 
standard $\gamma \gamma \to e^+ e^-$ process and thereby disappearing. Denoting 
by $E$ and $\epsilon$ the energy of the hard (incident) and of the soft (background) 
photon, respectively, and by $\varphi$ the scattering angle, the corresponding cross 
section is~(Heitler 1960)
\begin{eqnarray}
\lefteqn{
\sigma_{\gamma \gamma}(E, \epsilon, \varphi) ~=~ 1.25 \times 10^{-25} ~(1-\beta^2)~~\times }
	\nonumber\\
 & & ~~~~~~
\times ~~ 2\,\beta\,(\beta^2-2) ~+~ (3-\beta^4) ~{\rm ln} {1+\beta \over 1-\beta}~~
{\rm cm}^2 \, , 
\label{eq:sez.urto}
\end{eqnarray}
which depends on $E$, $\epsilon$ and $\varphi$  through the dimensionless parameter\footnote{Natural 
Lorentz-Heaviside units with $\hbar=c= K_{\rm B}=1$ are employed throughout.} 
$\beta(E,\epsilon,\varphi) \equiv \left[ 1 - 2 m_e^2/E\,\epsilon \left(1-\cos \varphi 
\right) \right]^{1/2}$. This process is kinematically allowed for $\epsilon$ above the 
energy threshold ${\epsilon}_{\rm thr}(E,\varphi) \equiv 2 m_e^2/ E \left(1-\cos \varphi 
\right)$, i.e. for $\beta > 0$. $\sigma_{\gamma \gamma}(E,\epsilon,\varphi)$ reaches its 
maximum, ${\sigma}_{\gamma \gamma}^{\rm max} \simeq 1.70 \cdot 10^{- 25} \, {\rm cm}^2$, 
for $\beta \simeq 0.70$. Assuming head-on collisions, ${\sigma}_{\gamma \gamma}^{\rm max}$ 
is attained when the background photon energy is $\epsilon_* (E) \simeq \left(0.5 \, {\rm 
TeV}/E \right) \, {\rm eV}$. This shows that in the energy interval $100 \, {\rm GeV} < E < 100 \, 
{\rm TeV}$, explored by the IACTs, the resulting opacity is dominated by the interaction with 
infrared/optical/ultraviolet diffuse background photons -- usually called Extragalactic 
Background Light (EBL) -- with $0.005 \, {\rm eV} < \epsilon < 5\, {\rm eV}$ (corresponding to the 
wavelength range $0.25 \, {\mu}{\rm m} < \lambda  < 250 \, {\mu}{\rm m}$).

The EBL is produced during the star-formation history of the Universe, including a possible 
early generation of stars formed before galaxies were assembled. Based on synthetic models of 
the evolving stellar populations in galaxies as well as on deep galaxy counts~(see, for a review, 
Hauser \& Dwek 2001), several estimates of the spectral energy distribution (SED) of the EBL have 
been proposed, leading to different values for the transparency of the Universe to $100 \, {\rm GeV} 
< E < 100 \, {\rm TeV}$ photons (Stecker et al. 1992 and 2006; Kneiske et al. 2002 and 2004; Primack et al. 
2005; Aharonian et al. 2006; Stanev et al. 2006; Mazin \& Goebel 2007; Franceschini et al. 2008); 
the resulting uncertainties are large. Besides including the evolution arising from the cosmic expansion, 
the evolutionary effects of the stellar populations of galaxies should be taken into account in the evaluation 
of the SED of the EBL (Raue \& Mazin 2008).

Because of the absorption produced by the EBL, the propagation of a monochromatic photon beam of 
observed energy $E_0$ from a source at redshift $z$ is controlled by the optical depth 
${\tau}_{\gamma}(E_0,z)$\footnote{We are adopting a standard $\Lambda$CDM cosmological model 
with ${\Omega}_{\Lambda} \simeq 0.7$ and ${\Omega}_{M} \simeq 0.3$. Distances are expressed in 
terms of the redshift $z$ and we shall henceforth write $E(z) = E_0 (1+z)$ and ${\epsilon}(z) = 
{\epsilon}_0 (1+z)$, with $E_0$ and ${\epsilon}_0$ referring to the present cosmic epoch ($z=0$).}. 
Therefore, the observed photon spectrum $\Phi_{\rm obs}(E_0,z)$ is related to the emitted one 
$\Phi_{\rm em}(E(z))$ by 
\begin{equation} 
\label{a0}
\Phi_{\rm obs}(E_0,z) = e^{- \tau_{\gamma}(E_0,z)} \ \Phi_{\rm em} \left( E_0 (1+z) \right)~. 
\end{equation}
Clearly ${\tau}_{\gamma}(E_0,z)$ increases with $z$, since a greater source distance entails 
a larger probability for a beam photon to be absorbed. It is generally assumed that $\Phi_{\rm em} 
\propto E^{- {\Gamma}_{\rm em}}$ for $E > 100 \, {\rm GeV}$, and so from Eq. (\ref{a0}) it follows that 
$\Phi_{\rm obs}(E_0,z)$ decreases exponentially with $z$. From now on, we restrict our attention 
to the energy range $0.2 \, {\rm TeV} < E_0 < 2 \, {\rm TeV}$ in which the blazar spectra have been measured 
by IACTs. In this range, data are fitted as $\Phi_{\rm obs} \propto E^{- {\Gamma}_{\rm obs}}$. Hence, 
Eq. (\ref{a0}) yields
\begin{equation}
\label{lunghexplZ}
{\Gamma}_{\rm obs}(z) \sim {\Gamma}_{\rm em} + \tau_{\gamma}(E_0,z)~,
\end{equation}
up to a logarithmic $z$-dependence. The experimental results are summarized in Table 1 and the spectral slopes 
are plotted vs. redshift in Fig.1.

\begin{table}
\caption{\label{fig:AGNtable}
Blazars with known redshift and VHE $\gamma$-ray flux and spectrum.}
\begin{tabular}{ l l l l }
\hline
\noalign{\smallskip}
Source    & $z$ & ${\Gamma}_{\rm obs}$ & $\Phi_{\rm obs}(> 0.2 \, {\rm TeV})$ \\
\noalign{\smallskip}
\hline
\noalign{\smallskip}
Mrk\,421       &  0.031 & $2.33 \pm 0.08$ & (1.0$\pm$0.1)$\times10^{-10}$ \\
Mrk\,501       &  0.034 & $2.28 \pm 0.05$ & (1.7$\pm$0.1)$\times10^{-11}$ \\
               &           & $2.45 \pm 0.07$ & (3.8$\pm$1.0)$\times10^{-12}$ \\
1ES\,2344+514  &  0.044 & $2.95 \pm 0.12$ & (1.2$\pm$0.1)$\times10^{-11}$ \\
Mrk\,180       &  0.045 & $3.30 \pm 0.70$ & (8.5$\pm$3.4)$\times10^{-12}$ \\
1ES\,1959+650  &  0.047 & $2.72 \pm 0.14$ & (3.0$\pm$0.4)$\times10^{-11}$ \\
BL\,Lacertae   &  0.069 & $3.60 \pm 0.50$ & (3.3$\pm$0.3)$\times10^{-12}$ \\
PKS 0548-322   &  0.069 & $2.80 \pm 0.30$ &   (3.3$\pm$0.7)$\times10^{-12}$ \\
PKS\,2005-489  &  0.071 & $4.00 \pm 0.40$ & (3.3$\pm$0.5)$\times10^{-12}$ \\
RGB\,J0152+017 &  0.080 & $2.95 \pm 0.36$ & (4.4$\pm$1.2)$\times10^{-12}$ \\
PKS\,2155-304  &  0.116 & $3.37 \pm 0.07$ & (2.9$\pm$0.2)$\times10^{-11}$ \\
1ES\,1426+428  &  0.129 & $3.55 \pm 0.46$ & (2.5$\pm$0.4)$\times10^{-11}$ \\
1ES\,0229+200  &  0.139 & $2.50 \pm 0.19$ & (4.5$\pm$0.7)$\times10^{-12}$ \\
H\,2356-309    &  0.165 & $3.09 \pm 0.24$ & (2.5$\pm$0.7)$\times10^{-12}$ \\
1ES\,1218+304  &  0.182 & $3.00 \pm 0.40$ & (1.0$\pm$0.3)$\times10^{-11}$ \\
1ES\,1101-232  &  0.186 & $2.94 \pm 0.20$ & (4.4$\pm$0.7)$\times10^{-12}$ \\
1ES\,0347-121  &  0.188 & $3.10 \pm 0.23$ & (3.9$\pm$0.7)$\times10^{-12}$ \\
1ES\,1011+496  &  0.212 & $4.00 \pm 0.50$ & (6.4$\pm$0.3)$\times10^{-12}$ \\
PG\,1553+113   &  $>$0.25& $4.20 \pm 0.30$ & (5.2$\pm$0.9)$\times10^{-12}$ \\
3C\,279        &  0.536 & $4.1 \pm 0.7$  &   (2.9$\pm$0.5)$\times10^{-11}$ \\
\noalign{\smallskip}
\hline
\end{tabular}
\noindent
The observed photon spectral index ${\Gamma}_{\rm obs}$ in the $0.2 \, {\rm TeV} 
< E_0 < 2 \, {\rm TeV}$ band ($0.2 \, {\rm TeV} < E_0 < 0.6 \, {\rm TeV}$ for 3C\,279: 
the effective $0.2 \, {\rm TeV} < E_0 < 2 \, {\rm TeV}$ slope of this source might be 
steeper) and the flux $\Phi_{\rm obs} (> 0.2 \, {\rm TeV})$ measured at energy 
$E_0 > 0.2 \, {\rm TeV}$ (in ${\rm erg} \, {\rm cm}^{-2} \, {\rm s^{-1}}$), for blazars 
observed at different redshift $z$. The errors indicate the statistical uncertainty; the 
corresponding systematic uncertainties on the spectral index are typically $\sim$0.1 
for H.E.S.S.\ and $\sim$0.2 for MAGIC. See De Angelis, Mansutti \& Persic (2008) 
for references and more informations.
\end{table}

A full-fledged prediction of how a broadband ($100 \, {\rm GeV} < E_0 < 100 \, {\rm TeV}$) 
SED of blazars is affected by the EBL would be beyond the scope of this paper. We will, instead, 
examine modifications of blazar spectra due to physical processes possibly taking 
place in intergalactic space and discuss them with respect to current IACT data. 
Such data cover the $0.2 \, {\rm TeV} < E_0 < 2 \, {\rm TeV}$ energy range, hence the corresponding 
relevant EBL range is $0.25 \, {\rm eV} < {\epsilon}_0 < 2.5 \, {\rm eV}$. Within this range, we find it 
convenient to adopt the following analytic parametrization of the EBL spectral number density at the 
present cosmic epoch (Stecker et al. 1992, SDS)
\begin{equation}
\label{pndq1x}
n_{\gamma}(\epsilon_0, 0) \simeq 10^{-3} \, \alpha  \left(\frac{\epsilon_0}{{\rm eV}} \right)^{- 2.55} \, 
{\rm cm}^{-3} \, {\rm eV}^{-1}~,
\end{equation}
where $\alpha$ is a suitable constant. Indeed, in the range $0.25 \, {\rm eV} < {\epsilon}_0 < 2.5 \, {\rm eV}$, 
Eq. (\ref{pndq1x}) does reproduce the basic behavior of the SED reported in the most 
recent phenomenological model of the EBL (Franceschini et al. 2008) for $0.5 < {\alpha} < 3$\footnote{The 
SED in the model of Franceschini et al. (2008) is slightly convex in the ${\rm log} \, \left[\epsilon_0 \, 
n_{\gamma}({\epsilon}_0, 0) \right]$ vs. ${\rm log} \, {\epsilon}_0$ representation for $0.25 \, {\rm eV} < 
\epsilon_0 < 2.5 \, {\rm eV}$, due to the emission bump resulting from the integrated emission of the low-mass 
star population that remains close to the main sequence over cosmological times (see e.g. Kneiske et al. 2002). 
The values $\alpha = 0.5$ and $\alpha = 3$ bracket a linear stripe in the  ${\rm log} \, \left[\epsilon_0 \, 
n_{\gamma}({\epsilon}_0, 0) \right]$-${\rm log} \, {\epsilon}_0$ plane which envelopes the actual EBL shape.}.
We remark that in the band $0.25 \, {\rm eV} < {\epsilon}_0 < 2.5 \, {\rm eV}$ the assumed SDS parametrization also 
encompasses both the lowest EBL level predicted by Primack et al. (2005) and the higher EBL level predicted by 
the ``baseline model'' of Stecker et al. (2006)\footnote{In this paper we disregard the ``fast evolution model'' of 
Stecker et al. (2006), which predicts an even higher EBL level and would require $\alpha > 3$.}. 

From the definition of $\tau_{\gamma}(E_0, z)$~(see e.g. Fazio \& Stecker 1970), we finally obtain
\begin{equation}
\label{eq:taubisq}
\tau_{\gamma}(E_0, z) \simeq 0.50 \, \alpha \,  \left( \frac{E_0}{500 \, {\rm GeV}} \right)^{1.55} \, 
\left[ \left( 1 + z \right)^{4.4} - 1 \right]~,
\end{equation}
where evolutionary effects arising from galaxy evolution (Raue \& Mazin 2008) have been included on top 
of those produced by the cosmic expansion. We have checked that for $1.2 < {\alpha} < 3$, $0.2 \, {\rm eV} < 
{\epsilon}_0 < 2.5 \, {\rm eV}$ and $0.05 < z < 0.5$ Eq. (\ref{eq:taubisq}) is approximately consistent with the 
analytic fit, provided by Stecker \& Scully (2006), of $\tau_{\gamma}(E_0, z)$ as predicted by the `baseline 
model' of Stecker et al. (2006).

Since for the nearby blazars ($z < 0.03$) the EBL photon absorption is expected to be negligible, 
we assume that for such sources observations do yield $\Phi_{\rm em}$. We find, on average, 
${\Gamma}_{\rm em} \simeq 2.4$
	\footnote{Several interpretations for the scattering of the data in Fig. 1 
		for $z < 0.2$ are possible: e.g., different sources can be observed in 
		different emission states, and so they may exhibit slightly different 
		slopes~(Persic \& De Angelis 2008).}. 
Various emission models for blazars have been developed and are briefly summarized in the Appendix. In the 
widely used Synchro-Self-Compton (SSC) emission model (e.g., Ghisellini et al. 1998) ${\Gamma}_{\rm em} \simeq 
2.4$ suggests a Compton peak at around or below $100 \, {\rm GeV}$. The predicted observed spectral index 
${\Gamma}_{\rm obs}^{\rm SDS}$ then follows from Eqs. (\ref{lunghexplZ}) and (\ref{eq:taubisq}): it is 
represented as the light-grey area in Fig. 1. 
%
\begin{figure}
\vspace{6cm}
\includegraphics{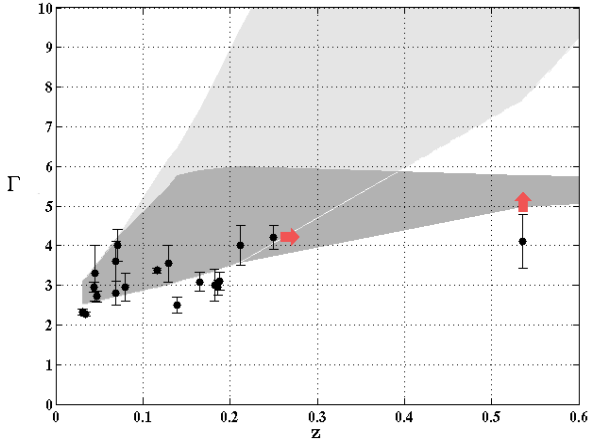}
\caption{Observed values of the spectral index for all blazars detected so far in the VHE band 
are represented by dots and corresponding error bars. Superimposed on them is the predicted 
behaviour of the observed spectral index within two different scenarios. In the first scenario (light grey area)
${\Gamma}_{\rm obs}^{\rm SDS}$ is computed in terms of standard physics in the SDS model of the EBL. In the DARMA 
scenario (dark grey area) ${\Gamma}_{\rm obs}^{\rm DARMA}$ is evaluated within the proposed photon-ALP 
oscillation mechanism as based on the same SDS model of the EBL.}
\label{landfig}
\end{figure}
%

Fig. 1 shows that the actually observed spectral index increases more slowly than ${\Gamma}_{\rm 
obs}^{\rm SDS}$ for redshifts $z > 0.2$. Moreover, the observed values cannot be explained for $z 
> 0.3$ by the EBL model of SDS even for $\alpha$ as low as 0.5. Being the optical depth a 
monotonically increasing function of $z$, we interpret the conflict between ${\Gamma}_{\rm obs}$ 
and ${\Gamma}_{\rm obs}^{\rm SDS}$ -- apparent in Fig. 1 -- as calling for a departure from the conventional 
view and we proceed to explore its consequences.

A possible way out relies upon a systematic hardening of the emission spectrum with increasing $z$. 
That is, ${\Gamma}_{\rm em}$ -- which is currently supposed to be independent of $z$ -- has to 
decrease as $z$ increases, so as to offset the growth of $\tau_{\gamma}(E_0, z)$. 
This situation is very difficult to achieve within the SSC emission model: even assuming that we 
selectively observe increasingly flaring sources at higher redshifts, the radiating electrons will 
be emitting more and more in the Klein-Nishina regime. Therefore -- unlike the synchrotron peak 
which will appreciably shift to higher energies -- the Compton peak will hardly shift, thereby 
ensuring that ${\Gamma}_{\rm em}$ is indeed virtually independent of $z$~(Persic \& De Angelis 2008). 

Other possibilities have been suggested, based on modifications of the emission 
mechanism. One involves strong relativistic shocks, which can give rise to values of ${\Gamma}_{\rm 
em}$ considerably smaller than previously thought~(Stecker et al. 2007; Stecker \& Scully 2008). 
Another rests upon photon absorption inside the blazar, which has be shown to lead to a substantial 
change of ${\Gamma}_{\rm em}$~(Aharonian et al. 2008a). A further option could be the 
underlying emission mechanism of the more luminous sources (e.g., flat-spectrum radio quasars) 
being hadronic, with the muon (and cascade) synchrotron component peaking in the sub-TeV region and 
the (neutral- and charged-)pion cascades crossing the 0.2-2 TeV band with a very hard ($\Gamma \sim 1.9$) 
spectrum, as shown in the very case of 3C\,279 (B\"ottcher et al. 2008).  

While successful at substantially reducing ${\Gamma}_{\rm em}$ in individual sources, these 
attempts fail to provide a systematic in the correlation of ${\Gamma}_{\rm em}$ versus $z$, needed 
to overcome the above difficulty. 

In the framework of the same basic emission mechanism being at work in blazars and flat-spectrum 
quasars both at low and at high redshift, we propose a different solution, referred to as the DARMA 
scenario~(De Angelis, Roncadelli \& 
Mansutti 2007). Implicit in all previous considerations is the hypothesis that photons propagate in the 
standard way throughout cosmic distances. We suppose instead that, in the presence of cosmic magnetic 
fields, photons can oscillate into a new, very light, spin-zero particle -- named Axion-Like Particle 
(ALP) -- and vice-versa. Once ALPs are produced close enough to the source, they travel unimpeded 
throughout the Universe and can convert back to photons before reaching the Earth. Since ALPs do not 
undergo EBL absorption, they act as if the observed photons had an effective optical depth smaller than 
$\tau_{\gamma}(E_0, z)$. Eq. (\ref{lunghexplZ}) entails that ${\Gamma}_{\rm obs}$ gets reduced by the 
same amount as far as the $z$-dependence is concerned, thereby avoiding the 
conflict shown in Fig. 1. Now the dependence of ${\Gamma}_{\rm obs}$ on $z$ is in agreement with 
observations, since the photon-ALP oscillation reduces photon absorption even for standard emission 
spectra. In order to guarantee consistency with observations of nearby blazars we take ${\Gamma}_{\rm 
em} \simeq 2.4$ for all sources represented in Fig. 1.

The key ingredient of the DARMA scenario -- namely the existence of ALPs -- is not an {\it ad hoc} 
assumption invented to solve the problem in question\footnote{Other aspects concerning the relevance 
of ALPs for gamma-ray astrophysics have been addressed in~(Hooper \& Serpico 2007; Hochmuth \& Sigl 
2007; De Angelis, Mansutti \& Roncadelli 2008; Simet et al. 2008; Cheleuche et al. 2008, Cheleuche \& 
Guendelman 2008).}. Instead, 
very light ALPs turn out to be a generic prediction of many extensions of the Standard Model of 
elementary particle physics and have attracted considerable interest over the past few years. Besides 
than in four-dimensional models~(Masso \& Toldra 1995 and 1997; Coriano \& Irges 2007; Coriano et al. 2007), 
they naturally arise in the context of compactified Kaluza-Klein theories~(Chang et al. 2000; Dienes 
et al. 2000) as well as in superstring theories~(Turok 1996; Svrcek \& Witten 2006). Moreover, it has 
been argued that an ALP with mass $m \sim 10^{-33}\,{\rm eV}$ is a good candidate for the 
quintessential dark energy~(Carroll 1998) which might trigger the present accelerated cosmic expansion.

Below, we first review the motivation and the properties of ALPs which are of direct relevance for the 
present discussion. Next, we outline the computation of the predicted observed spectral indices 
${\Gamma}_{\rm obs}^{\rm DARMA}$. Details on the derivation, as well as on the dependence of 
${\Gamma}_{\rm obs}^{\rm DARMA}$ on the adopted EBL model, will be reported elsewhere.

On the basis of phenomenological as well as conceptual arguments, the Standard Model is currently viewed 
as the low-energy manifestation of some more fundamental and richer theory of all elementary-particle 
interactions including gravity. Therefore, the lagrangian of the Standard Model is expected to be modified 
by small terms describing interactions among known and new particles. ALPs are spin-zero light bosons 
defined by the following low-energy effective lagrangian
\begin{equation}
\label{a1a}
{\cal L}_{\rm ALP} \ = \ 
\frac{1}{2} \, \partial^{\mu} \, a \, \partial_{\mu} \, a - \frac{1}{2} 
\, m^2 \, a^2 - \frac{1}{4 M} \, F^{\mu \nu} \, \tilde F_{\mu \nu} \, a~,
\end{equation}
where $F^{\mu \nu}$ is the electromagnetic field strength, $\tilde F_{\mu \nu}$ is its dual, 
$a$ denotes the ALP field whereas $m$ stands for the ALP mass. According to the above view, 
it is assumed for the inverse two-photon coupling $M \gg 
G_F^{- 1/2} \simeq 250 \, {\rm GeV}$. On the other hand, it is supposed that $m \ll G_F^{- 1/2} 
\simeq 250 \, {\rm GeV}$ and for definiteness we take $m < 1 \, {\rm eV}$. As far as generic 
ALPs are concerned, the parameters $M$ and $m$ are to be regarded as independent.

So, what really characterizes ALPs is the trilinear $\gamma$-$\gamma$-$a$ vertex described 
by the last term in ${\cal L}_{\rm ALP}$, whereby one ALP couples to two photons. Owing to 
this vertex, ALPs can be emitted by astronomical objects of various kinds, and the present 
situation can be summarized as follows. The negative result of the CAST experiment designed 
to detect ALPs emitted by the Sun yields the bound $M > 0.86 \cdot 10^{10} \, {\rm GeV}$ for 
$m < 0.02 \, {\rm eV}$~(Zioutas et al. 2005; Andriamoje et al. 2007). Moreover, theoretical 
considerations concerning star cooling via ALP emission provide the generic bound $M > 10^{10} 
\, {\rm GeV}$, which for $m < 10^{- 10} \, {\rm eV}$ gets replaced by the stronger one 
$M >  10^{11} \, {\rm GeV}$  even if with a large uncertainty~(Raffelt 1990 and 1996; Khlopov et 
al. 2004).

The same $\gamma$-$\gamma$-$a$ vertex produces an off-diagonal element in the mass matrix for 
the photon-ALP system in the presence of an external magnetic field ${\bf B}$. Therefore, the 
interaction eigenstates differ from the propagation eigenstates and photon-ALP oscillations show 
up~(Sikivie 1984a and 1984b; Maiani et al. 1986; Raffelt \& Stodolsky 1988). The situation is 
analogous to what happens in the case of massive neutrinos with different flavours, but while 
all neutrinos have equal spin -- hence neutrino oscillations can freely occur -- ALPs have 
instead spin zero whereas the photon has spin one, and so the transformation can take place only 
if the spin mismatch is compensated for by an external magnetic field ${\bf B}$. 

We imagine that a sizeable fraction of photons emitted by a blazar soon convert into ALPs. They 
propagate unaffected by the EBL and we suppose that before reaching the Earth a substantial fraction 
of ALPs convert back into photons. We further assume that this photon-ALP oscillation process 
is triggered by cosmic magnetic fields, whose existence has been demonstrated very recently by AUGER 
observations~(Abraham et al. 2007). Lacking information about their morphology, we assume that cosmic 
magnetic fields have a domain-like structure~(Kronberg 1994; Grasso \& Rubinstein 
2001; Furlanetto \& Loeb 2001): ${\bf B}$ is supposed to be constant over a domain of 
size $L_{\rm dom}$ equal to its coherence length and to change randomly its direction from one 
domain to another while keeping approximately the same strength. As argued by De Angelis, 
Persic \& Roncadelli (2008), it looks plausible to assume the coherence length in the range 
$1 \, {\rm Mpc} < L_{\rm dom} < 10 \, {\rm Mpc}$. Correspondingly, the inferred strength lies in the range 
$0.3 \, {\rm nG} < B_0 <  0.9 \, {\rm nG}$. This conclusion agrees with previous upper bounds~(Blasi et al. 1999), 
so we assume $L_{\rm dom} = 7 \, {\rm Mpc}$ and $B_0 = 0.5\, {\rm nG}$ as reference values at $z = 0$. 

Following the same computational procedure as in De Angelis, Roncadelli \& Mansutti (2007), we evaluate 
the probability $P_{\gamma \to \gamma}(E_0,z)$ that a photon remains a photon after propagation from 
the source to us when allowance is made for photon-ALP oscillations as well as for photon absorption by 
the EBL. As a consequence, Eq. (\ref{a0}) becomes
\begin{equation}
\label{a0bis}
\Phi_{\rm obs}(E_0,z) = P_{\gamma \to \gamma}(E_0,z) \ \Phi_{\rm em} \left( E_0 (1+z) \right) 
\end{equation}
so that Eq. (\ref{lunghexplZ}) gets replaced by
\begin{equation}
\label{lunghexplZW}
{\Gamma}_{\rm obs}(z) \sim {\Gamma}_{\rm em} - {\rm ln} \, P_{\gamma \to \gamma}(E_0,z)~,
\end{equation}
again up to a logarithmic $z$-dependence. Assuming $m < 10^{- 10} \, {\rm eV}$ and $M \simeq 4 \cdot 
10^{11} \, {\rm GeV}$ as in De Angelis, Roncadelli \& Mansutti (2007) and adopting the same EBL model 
as before (see Eq. (\ref{pndq1x}) with $0.5 < {\alpha} < 3$), our result for ${\Gamma}_{\rm obs}^{\rm DARMA}$ 
is represented by the dark gray area in Fig. 1. We have checked that the same result remains practically 
unaffected within the range $10^{11} \, {\rm GeV} < M < 10^{13} \, {\rm GeV}$. 

In conclusion, for a realistic EBL model (defined by Eq. (\ref{pndq1x}) with $0.5 < {\alpha} < 3$) and assuming the 
same nominal emission spectral slope ${\Gamma}_{\rm em} \simeq 2.4$ for all VHE blazars, the DARMA scenario 
naturally explains the IACT data and predicts that ${\Gamma}_{\rm obs}$ becomes asymptotically independent of 
$z$ for far-away sources. Our prediction can be tested with the satellite-borne {\it Fermi}/LAT detector as well as 
with the ground-based Cherenkov telescopes H.E.S.S., MAGIC, CANGAROO\,III, VERITAS and the Extensive Air Shower 
arrays ARGO-YBJ and MILAGRO. We remark that the DARMA scenario could loose much of its motivation 
-- and be eventually disproved -- if the emission mechanisms of VHE blazars and quasars had a variation according 
to, e.g., luminosity. The most distant and luminous VHE $\gamma$-ray source that appears in Fig.1 is 3C\,279, 
a remarkable example of flat-spectrum radio galaxies: for these sources, the flaring and accompanying intermittency 
of source activity, not well understood at present, may point to emission mechanisms different from those that are 
commonly being used for blazars (i.e., the leptonic SC models). Such emission mechanisms may provide the hard spectra 
emitted by high-luminosity, high-$z$ sources, which in Fig.1 are required to counterbalance the spectral 
steepening imposed to TeV radiation by traversing the EBL over cosmological distances.

\section*{Acknowledgments}

M.R. thanks the Dipartimento di Fisica Nucleare e Teorica, Universit\`a di Pavia, for support. 
We thank an anonymous referee for useful suggestions.

\section*{Appendix}

The observed emission from a blazar is mainly interpreted as radiation from a relativistic jet that is 
directed approximately along the line of sight to the observer (Urry \& Padovani 1995). 

Leptonic models associate 
the TeV emission with the synchrotron-Compton (SC) process occurring in the jet. In one/more magnetized plasma 
blob(s) moving relativistically toward the observer, the electrons emit synchrotron radiation that is, in turn, Compton 
upscattered by their parent electrons (one/multi-zone synchro-self-compton [SSC] scheme: e.g., Maraschi et al. 1992). 
A variant of this scheme involves an additional, external field of soft photons  -- e.g., the infrared field associated with 
the broad-line region (external inverse Compton [EIC] scheme: Dermer et al. 1992). Such leptonic models are mainly 
motivated by the double-hump shape generally observed in blazar SEDs. However, looseness/lack of X-ray/TeV variability 
correlation including 'orphan' TeV flares (e.g., Krawczynski et al. 2004; Blazejowski et al. 2006), and peculiar quasar SEDs 
(e.g., B\"ottcher et al. 2008) increasingly challenge the simple -- hence popular --  SC model. 

Hadronic processes are more 
complicated: they can entail photon-initiated cascades (Mannheim \& Biermann 1992) or pp collisions (Pohl \& Schlickeiser 2000; 
Dar \& Laor 1997; Beall \& Bednarek 1999), but both essentially produce VHE photons via $\pi^0$ decay. For TeV blazars it 
was suggested that synchrotron emission from ultra-relativistic protons in strong magnetic fields is mostly responsible for the 
observed TeV emission from blazars (Aharonian 2000; M\"ucke et al. 2003). The hadronic model may successfully describe 
the broad-band SED of blazars (Aharonian et al. 2005; B\"ottcher et al. 2008) and, in principle, account for the X-ray/TeV 
correlation, if an appreciable amount of X-ray emission comes from the synchrotron radiation of secondary electrons. There is in 
fact some flexibility here, as the co-accelerated, primary electrons might also contribute to the X-ray band in a significant 
manner. This possibly 'double' origin of the X-rays may explain the scatter in the X-ray/TeV correlation. 

Current leptonic and hadronic models, however, are challenged by the 'orphan' flares and by the minute/hour-scale variability 
of blazars both at TeV energies (Gaidos et al. 1996) and X-ray energies (Cui 2004; Xue \& Cui 2005). The flaring and 
accompanying intermittency of source activity is probably a key to understanding the physics of flat-spectrum radio sources, of 
which the outlier in Fig.1, 3C\,279, is a remarkable example. Furthermore, if ultra-high-energy ($E \magcir 10^{20} \, {\rm eV}$) 
cosmic rays (Linsley 1963; Protheroe \& Clay 2004) are produced in AGNs, then purely leptonic models are certainly limited and -- 
eventually -- inadequate. For general reviews on the origin of astrophysical TeV photons, see e.g. Cui (2006), Aharonian 
et al. (2008b), and De\,Angelis, Mansutti \& Persic (2008), and references therein.

\end{document}